\documentclass[letterpaper]{article} 
\usepackage{aaai24}  
\usepackage{times}  
\usepackage{helvet}  
\usepackage{courier}  
\usepackage[hyphens]{url}  
\usepackage{graphicx} 
\usepackage{natbib}  
\usepackage{caption} 
\usepackage{algorithm}
\usepackage{algorithmic}

\usepackage{subfigure}
\usepackage{amsmath}
\usepackage{amssymb}
\usepackage{xcolor}
\usepackage{colortbl}
\usepackage{tabularx}
\usepackage{booktabs}
\usepackage{multirow}
\usepackage{natbib}
\definecolor{mygray}{RGB}{230,230,230}
\usepackage[colorlinks=true, linkcolor=black,urlcolor=black, citecolor=teal]{hyperref}

\usepackage{newfloat}
\usepackage{listings}
\DeclareCaptionStyle{ruled}{labelfont=normalfont,labelsep=colon,strut=off} 
\floatstyle{ruled}
\newfloat{listing}{tb}{lst}{}
\floatname{listing}{Listing}
\title{\centering HiHPQ: Hierarchical Hyperbolic Product Quantization \\ for Unsupervised Image Retrieval}
\author{Zexuan Qiu, Jiahong Liu, Yankai Chen, Irwin King}
\affiliations{The Chinese University of Hong Kong\\
                \{zxqiu22, jhliu22, ykchen, king\}@cse.cuhk.edu.hk}

\usepackage{bibentry}

\begin{document}

\maketitle

\begin{abstract}
Existing unsupervised deep product quantization methods primarily aim for the increased similarity between different views of the identical image, whereas the delicate multi-level semantic similarities preserved between images are overlooked. Moreover, these methods predominantly focus on the Euclidean space for computational convenience, compromising their ability to map the multi-level semantic relationships between images effectively. To mitigate these shortcomings, we propose a novel unsupervised product quantization method dubbed \textbf{Hi}erarchical \textbf{H}yperbolic \textbf{P}roduct \textbf{Q}uantization (HiHPQ),  which learns quantized representations by incorporating hierarchical semantic similarity within hyperbolic geometry. Specifically, we propose a hyperbolic product quantizer, where the hyperbolic codebook attention mechanism and the quantized contrastive learning on the hyperbolic product manifold are introduced to expedite quantization. Furthermore, we propose a hierarchical semantics learning module, designed to enhance the distinction between similar and non-matching images for a query by utilizing the extracted hierarchical semantics as an additional training supervision. Experiments on benchmarks show that our proposed method outperforms state-of-the-art baselines.~\footnote{Our code is available at \url{https://github.com/zexuanqiu/HiHPQ}.}
 
\end{abstract}

\section{Introduction}

Approximate Nearest Neighbor (ANN) search has attracted considerable interest in contemporary image retrieval systems owing to its remarkable search efficiency and exceptional performance. There are two main branches in ANN research: \textit{Binary Hashing} (BH)~\cite{salakhutdinov2009semantic} and \textit {Product Quantization} (PQ)~\cite{jegou2010product}. These two approaches both transform data into binary codes. BH maps high-dimensional data to the Hamming space, allowing fast distance calculations using bit-wise XOR operation. However, it has limitations on image similarity due to integer-only distance values. As for PQ, it decomposes the high-dimensional space and approximates distances between binary codes using pre-computed real-valued inter-codeword distance, reflecting richer similarity information. In industrial practice, PQ has been incorporated into numerous fast retrieval engines~\cite{avq_2020}.

\begin{figure}
    \centering
    \includegraphics[width=0.9\linewidth]{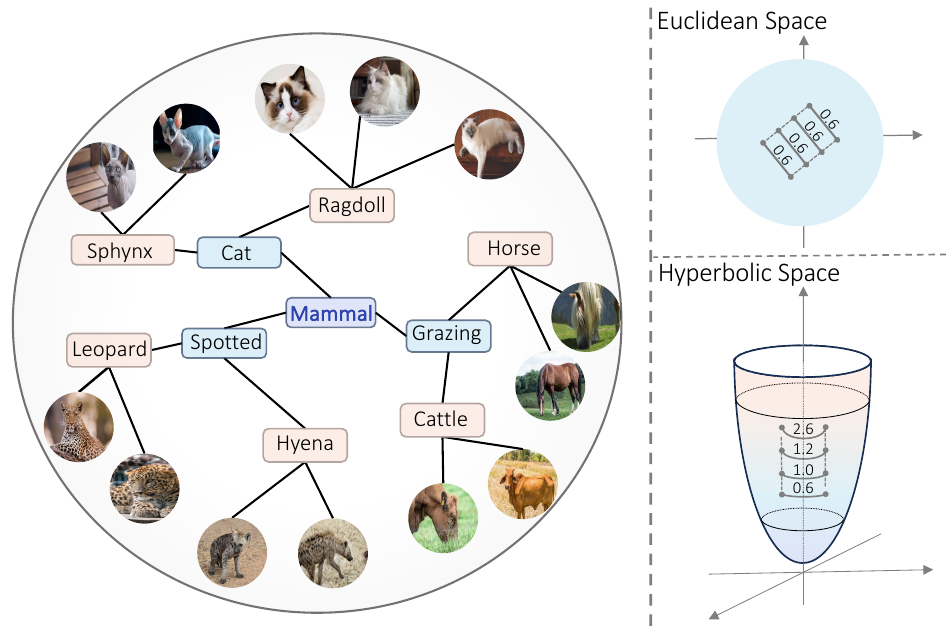}
    \vspace{-3mm}
    \caption{\textbf{Left:} Illustration of hierarchical semantics preserved in images. \textbf{Right:} Distance comparison on Euclidean space (top) and hyperbolic space (bottom). The $2$-dimensional hyperbolic space example is depicted using the Lorentz model $\mathcal{L}^2$ in $\mathbb{R}^3$.  }
    \label{fig: intro}
\vspace{-5mm}
\end{figure}

The standard PQ~\cite{jegou2010product} and its variants(\textit{e.g.}, OPQ~\cite{ge2013optimized} and LOPQ~\cite{kalantidis2014locally}) are widely used in large-scale image retrieval systems. They rely on well-learned representations and heuristic algorithms for training the quantization module. Deep supervised PQ methods~\cite{yue2016deep,yu2018product,liu2018deep,klein2019end} have been developed to learn the neural encoding network and quantization module together using annotated labels. However, their practical application is challenging due to the high cost of annotating a large number of images. To address this, limited unsupervised deep PQ methods have emerged, utilizing reconstruction-based~\cite{morozov2019unsupervised} and contrastive objectives~\cite{jang2021self,wang2022contrastive} to optimize quantized representations.

Current unsupervised product quantization methods based on contrastive learning solely focus on striving for greater similarity between different views of the same image, disregarding the inherent hierarchical semantic structures commonly found in images. For instance, as illustrated in Figure~\ref{fig: intro}, visual characteristics bring together \textit{leopards} and \textit{hyenas} into a \textit{spotted} cluster, allowing them to be categorized more coarsely as \textit{mammals} alongside similar animals like \textit{cats}. Thus, in order to distinguish between the closest neighbors of a query and non-matching items during the retrieval phase of PQ, it is vital to understand the nuanced levels of similarity among instances, instead of simply treating all non-query images as negatives. As supporting evidence of our motivation, such hierarchical semantic similarity has been shown to be helpful for performance improvement in other fields like visual representations~\cite{guo2022hcsc}, metric learning~\cite{yan2021unsupervised}, classification~\cite{zhang2022use, liu2022discovering, liu2022cspm, ma2023graph} and style transfer~\cite{li2020unsupervised}.

On the other hand, \textit{Euclidean space} has become a workhorse for existing PQ methods, where the Euclidean distance between quantized representations serves as a measure of similarity between images. 
However, \textit{hyperbolic space} (\textit{i.e.}, non-Euclidean space) has shown less distance distortion than Euclidean space when fitting hierarchical data, such as word embeddings~\cite{tifrea2018poincar}, graphs~\cite{chen2022modeling, liu2022enhancing} and image-text retrieval~\cite{desai2023hyperbolic}. This can be attributed to the exponentially-evolved distance measurement where samples far away from the origin present a larger distance, as demonstrated on the right side of Figure~\ref{fig: intro}.  Consequently, by integrating hyperbolic geometry into the quantization representations, there is great potential to enhance the capturing of the structured semantics of an image.

Motivated by the aforementioned limitations, we introduce a new unsupervised PQ technique, dubbed \textbf{Hi}erarchical \textbf{H}yperbolic \textbf{P}roduct \textbf{Q}uantization (HiHPQ). Generally, it enhances quantized representation learning by considering the hierarchical semantics within hyperbolic geometry:

\begin{itemize}
    \item We propose a novel hyperbolic product quantizer. In this framework, each low-dimensional subspace is represented as a Lorentzian manifold with varying curvature, where the data points and codewords are both embedded in the hyperbolic space. Then a novel hyperbolic codebook attention mechanism is proposed to seamlessly extend the differentiable quantization process into the hyperbolic space. Additionally, we design a quantized contrastive learning approach based on the hyperbolic distance for model optimization.
    \item We design a hierarchical semantics learning module. Specifically, in addition to vanilla view-augmented semantics as the training signal, we utilize hierarchical clustering to extract pseudo hierarchical semantics as extra supervision for product quantization. To ensure retrieval of similar terms for the query, we extract two types of positive samples: the closest prototype in each hierarchy and instances sharing the same closest prototype. Then the prototype-wise and instance-wise contrastive losses that incorporate these two additional positive samples are employed to inject the extracted hierarchical semantics into quantized representations.
    
    \item Our HiHPQ surpasses state-of-the-art baselines on benchmark image datasets, with empirical analyses confirming the effectiveness of each proposed component.

\end{itemize}

\section{Related Work}
\paragraph{Product Quantization}
Product quantization was initially proposed in the field of source coding \cite{gray1984vector} and later applied to ANN search \cite{jegou2010product}. In large-scale image retrieval systems, PQ is often combined with inverted index~\cite{jegou2011searching,baranchuk2018revisiting} or inverted multi-index~\cite{babenko2014inverted} to achieve faster similarity search. Typical variants of PQ, such as OPQ~\cite{ge2013optimized} and Cartesian K-Means~\cite{norouzi2013cartesian}, focus on computing an optimized rotation matrix to make each subspace orthogonal to each other. Based on OPQ, LOPQ~\cite{kalantidis2014locally} employs a locally optimized individual product quantizer per cell of the inverted index.  Afterwards, a series of end-to-end supervised deep PQ methods for image retrieval~\cite{yue2016deep,liu2018deep,klein2019end,xiao2021matching} are proposed with powerful representation capabilities of neural networks. To address the expensive data annotation problem, limited unsupervised PQ methods have been proposed. Among them, the earlier approach~\cite{chen2018learning,morozov2019unsupervised} is based on the auto-encoder, while later methods~\cite{jang2021self,wang2022contrastive} are based on the contrastive learning frameworks. Also, \citet{chen2020differentiable} and \citet{lu2023differentiable} study improved techniques of differentiable quantization, particularly when the codebook adheres to certain constraints.

\paragraph{Binary Hashing}
In recent years, there has been a surge of research dedicated to unsupervised binary hashing~\cite{chen2023bipartite,bgr,chen2022effective, zhang2020discrete,zhang2019doc2hash}. 
An established type of unsupervised hashing method is based on deep generative models. Some of them\cite{dai2017stochastic,shen2019unsupervised,shen2020auto,li2021deep} utilize the encoder-decoder architecture~\cite{kingma2013auto} which expects the binary code to recover the original input, while others of them~\cite{dizaji2018unsupervised,zieba2018bingan,song2018binary} utilize generative adversarial networks~\cite{goodfellow2014generative} to implicitly maximize the reconstruction likelihood through the discriminator.  Another typical category of hashing methods aims to preserve predefined similarity between images using binary codes. In this category, methods like \cite{yang2018semantic,yang2019distillhash,tu2020mls3rduh} leverage pre-trained continuous features to define pairwise similarities, while methods like~\cite{lin2016learning,luo2020cimon, QiuSOYC21,lin2022deep} involves utilizing random augmentations to establish pairs of images that share common characteristics. This type of similarity-preserving method is also widely used in hashing-based video search~\cite{li2019neighborhood,li2021structure}.
Further, \citet{ma2022improved} and \citet{wei2022hyperbolic} both propose to employ fine-grained semantic similarity to improve hashing.

\section{Preliminaries}

\begin{figure*}[!t]
\centering
\subfigure[]{
\includegraphics[width=0.61\linewidth]{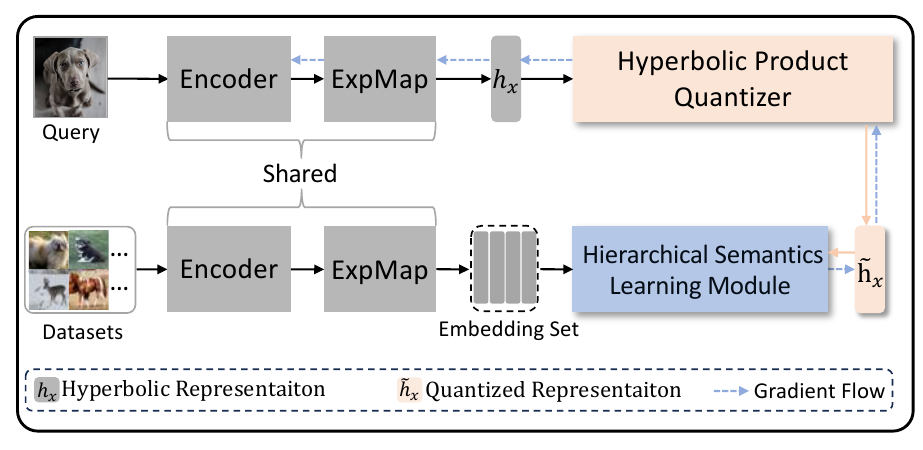}
\label{fig:framework}
}
\subfigure[]{
\includegraphics[width=0.36\linewidth]{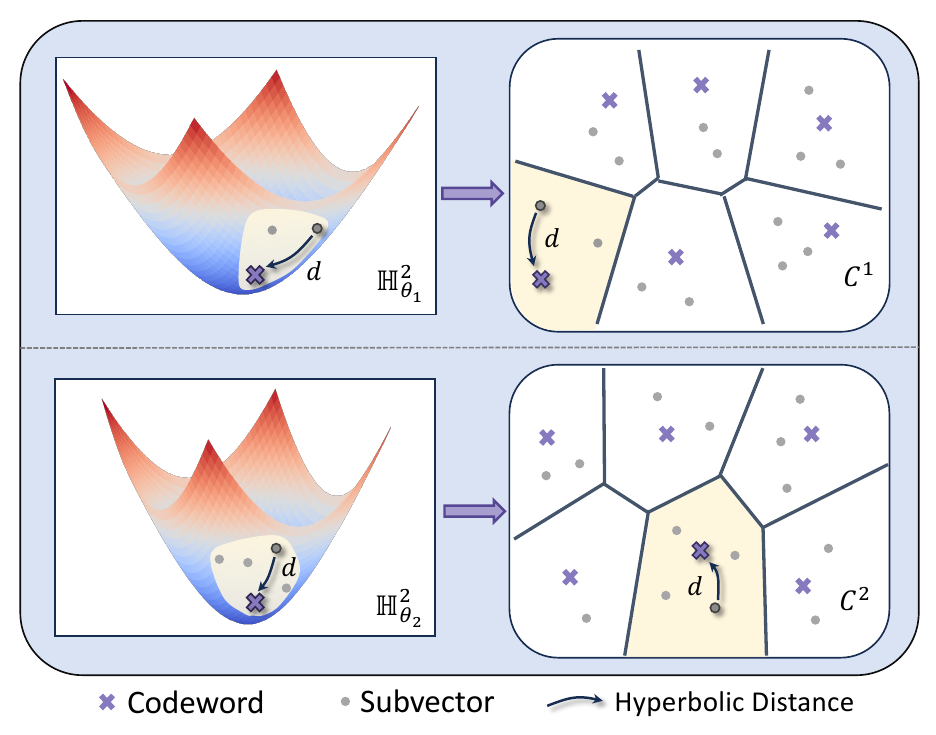}
\label{fig:hpq}
} 
\vspace{-3.5mm}
\caption{\textbf{(a)} The architecture of HiHPQ. ``ExpMap" denotes the exponential map for short. \textbf{(b)} An example of our hyperbolic product quantizer. In this example, there are two hyperbolic spaces $\mathbb{H}^2_{\theta_1}$ and $\mathbb{H}^2_{\theta_2}$ of 2-dimension depicted by the Lorentz model in $\mathbb{R}^3$, where $\theta_1$ and $\theta_2$ denotes the curvature parameter. On the right side, two codebooks $C^1$ and $C^2$  are displayed using 2D Voronoi diagrams. Subvectors will be quantized by codewords in corresponding codebooks via the hyperbolic distance metric.}
\vspace{-4mm}

\end{figure*}

\subsection{Lorentz Model of Hyperbolic Space}
In this part, we briefly review the concepts and basic operations on hyperbolic geometry. A thorough and in-depth explanation can be found in \cite{nickel2018learning}.
Multiple isometric models are commonly used to represent hyperbolic geometry. In this work, we adopt the typical Lorentz model to depict the hyperbolic space. 

\noindent{\textbf{Lorentz Model}}\ \ \ 
Lorentz model depicts the $d$-dimensional hyperbolic space (\textit{i.e.}, Lorentzian manifold) as a sub-manifold of $\mathbb{R}^{d+1}$. Formally, a $d$-dimensional Lorentz model with the constant negative curvature $-\theta$ is defined as the following set of vectors:
\begin{equation}
    \mathbb{H}^d_{\theta} = \{\mathbf{x} \in \mathbb{R}^{d+1}: \langle \mathbf{x},\mathbf{x} \rangle_\mathcal{L} = -1/\theta, \theta>0\},
\end{equation}
where $\langle \cdot, \cdot \rangle_\mathcal{L}$ denotes the \textit{Lorentzian inner product} of the hyperbolic space: $ \langle \mathbf{x}, \mathbf{y} \rangle _ \mathcal{L} = -x_0y_0 + x_1y_1 + x_2y_2 + \cdots x_dy_d.$

\noindent{\textbf{Lorentz Distance}}\ \ \ The Lorentzian distance between $\mathbf{x},\mathbf{y} \in \mathbb{H}^d_\theta$ is defined as:
\begin{equation}
    \mathrm{d}_{\mathbb{H}^d_{\theta}}(\mathbf{x},\mathbf{y}) = \sqrt{1/\theta} \ cosh^{-1}(- \theta \langle \mathbf{x}, \mathbf{y} \rangle_{\mathcal{L}}),
\label{fml:hyp_dist}
\end{equation}
which depicts the length of the shortest path (\textit{i.e.}, \textit{geodesic}) between two points on the manifold. With $\theta \rightarrow 0$, the distance in Eq.~\eqref{fml:hyp_dist} reduces the Euclidean distance.

\noindent{\textbf{Tangent Space}}\ \ \ For any vector $\mathbf{p} \in \mathbb{H}^d_{\theta}$, the tangent space  $\mathcal{T}_\mathbf{p} \mathbb{H}^d_{\theta}$ is defined as the first-order Euclidean approximation of  $\mathbb{H}^d_{\theta}$ around $\mathbf{p}$:
\begin{equation}
    \mathcal{T}_\mathbf{p} \mathbb{H}^d_{\theta} = \{\mathbf{v} \in \mathbb{R}^{d+1}: \langle \mathbf{v}, \mathbf{p} \rangle_\mathcal{L} = 0\}.
    \label{fml:tan}
\end{equation} 
For any ambient Euclidean vector $\mathbf{u} \in \mathbb{R}^{d+1}$, it can be projected onto the tangent space $\mathcal{T}_{\mathbf{p}} \mathbb{H}^d_{\theta}$ via the orthogonal projection $\text{proj}_{\mathbf{p}}(\mathbf{u})$:
\begin{equation}
    \text{proj}_{\mathbf{p}}(\mathbf{u}) = \mathbf{u} + \theta \mathbf{p} \langle \mathbf{p}, \mathbf{u} \rangle_\mathcal{L}.
    \label{fml:proj}
\end{equation}

\noindent{\textbf{Exponential Map}}\ \ \ The exponential map offers a method to project vectors from the tangent space onto the manifold. For a point $\mathbf{p}$ on the manifold, its exponential map is defined as  $\exp_\mathbf{p}^{\theta}:\mathcal{T}_\mathbf{p} \mathbb{H}^d_{\theta} \rightarrow \mathbb{H}^d_{\theta}$ is defined as:
\begin{equation}
    \exp_\mathbf{p}^{\theta}(\mathbf{v}) \! = \cosh(\sqrt{\theta} \left\|\mathbf{v} \right\| _{\mathcal{L}}) \mathbf{p} + \frac{\sinh(\sqrt{\theta} \left\| \mathbf{v} \right\| _\mathcal{L})}{\sqrt{\theta} \left\| \mathbf{v} \right\| _\mathcal{L}} \mathbf{v},
\end{equation}
where $\left\| \mathbf{v} \right\| _\mathcal{L} = \sqrt{\langle \mathbf{v}, \mathbf{v} \rangle_\mathcal{L}}$ is the Lorentzian norm of $\mathbf{v}$.

\section{Methodology}

\subsection{Problem Formulation and Model Overview}
The goal of our work is to learn a product quantization mapping: $x \mapsto \tilde{\mathbf{b}}_x$, where $x$ is an image of the unlabeled dataset $\mathcal{X}=\{x_i\}_{i=1}^N$, and $\tilde{\mathbf{b}}_x \in \{0,1\}^B$ is a $B$-bit quantization code. During the retrieval phase, $\tilde{\mathbf{b}}_x$ will recover the quantized representation $\tilde{\mathbf{h}}_x \in \mathbb{R}^D $ to enable efficient retrieval. Our proposed model, as illustrated in Figure~\ref{fig:framework}, is mainly composed of: (i) an encoder network that utilizes a (CNN) backbone followed by a linear projector, functioning as a feature extractor; (ii) a hyperbolic product quantizer endowed with  a proposed soft codebook quantization mechanism for quantizing representations in hyperbolic space; and (iii) a hierarchical semantics learning module designated to enhance the quantized representations based on the extracted hierarchical similarity.

\subsection{Hyperbolic Product Quantizer}

\paragraph{Hyperbolic Product Space}
Inspired by the recent development of hyperbolic manifold~\cite{gu2018learning,gao2022curvature}, instead of utilizing Euclidean geometry, we propose representing the vector space $\mathbb{S}$ as the Cartesian product of $M$ $d$-dimension hyperbolic subspaces:
\begin{equation}
    \mathbb{S} = \mathbb{H}^d_{\theta_1} \times \mathbb{H}^d_{\theta_2} \times \cdots \times \mathbb{H}^d_{\theta_M},
\end{equation}
where $-\theta_i(\theta_i > 0)$ denotes the negative constant curvature for the $i$-th Lorentzian manifold $\mathbb{H}^d_{\theta_i}$. Similarly, for the reference point $\mathbf{p}=(\mathbf{p}^1, \cdots, \mathbf{p}^M)$ on the manifold $\mathbb{S}$, the approximated tangent space $\mathcal{T}_{\mathbf{p}} \mathbb{S}$ of $\mathbf{p}$ can be decomposed by the Cartesian product of $M$ tangent spaces. Namely, for $\mathbf{v} \in \mathcal{T}_\mathbf{p} \mathbb{S}$,  we have $\mathbf{v}=(\mathbf{v}^1, \cdots, \mathbf{v}^M)$, where $\mathbf{v}^m \in \mathcal{T}_{\mathbf{p}^m} \mathbb{H}^d_{\theta_m}$.

During training, given an image $x$, we pass it through the encoder network to get the ambient Euclidean embedding $\mathbf{z}_x$ and slice it into $M$ segments $\{\mathbf{z}^m_x \}^M_{m=1}$.  Then, for each segment $\mathbf{z}^m_x \in \mathbb{R}^{d+1}$, we transform it onto the tangent space $\mathcal{T}_\mathbf{p}$ to get the tangent vector $\mathbf{v}^m_x$ via the orthogonal projection  $\mathbf{v}^m_\mathbf{x} = \text{proj}_{\mathbf{p}^m}(\mathbf{z}^m_\mathbf{x})$. For the obtained tangent vector $\mathbf{v}_x = (\mathbf{v}^1_x, \cdots, \mathbf{v}^M_x )$, we transfer it onto the product manifold $\mathbb{S}$ and get the final hyperbolic embedding $\mathbf{h}_x \in \mathbb{S}$ by:
\begin{equation}
    \mathbf{h}_x = \exp_{\mathbf{p}}(\mathbf{v}_x) = \left(\exp_{\mathbf{p}^1}^{\theta_1}(\mathbf{v}^1_x), \cdots, \exp_{\mathbf{p}^M}^{\theta_M}(\mathbf{v}^M_x) \right).
    \label{fml:exp}
\end{equation}
In practice, we adopt $\mathbf{p}=(\mathbf{o}^1,\cdots, \mathbf{o}^M)$ as the reference point when applying exponential map operations, where $\mathbf{o}^m = \left(\sqrt{1/\theta_m},\mathbf{0}\right) (\mathbf{0} \in \mathbb{R}^d)$ is the hyperbolic original point on the Lorentzian Manifold $\mathbb{H}_{\theta_m}^d$. 

As for the distance measurement on the product manifold $\mathbb{S}$, the distance between any pair $\mathbf{a}, \mathbf{b} \in \mathbb{S}$ simply decomposes into sum of $M$ Lorentzian distances:
\begin{equation}
    \mathrm{d}_{\mathbb{S}}(\mathbf{a}, \mathbf{b}) = \sum^M_{m=1}\mathrm{d}_{\mathcal{L}_m}(\mathbf{a}, \mathbf{b}),
\end{equation}
where $\mathrm{d}_{\mathcal{L}_m}(\mathbf{a}, \mathbf{b})$ denotes the Lorentzian distance of the $m$-th Lorentzian manifold $\mathbb{H}_{\theta_m}^d$.

\paragraph{Hyperbolic Codebook Quantization Mechanism}
The performance of product quantization is greatly influenced by the choice of the quantization technique. In recent methods~\cite{chen2020differentiable,jang2021self,wang2022contrastive}, the differentiable soft quantization technique has been widely used and proven more effective and feasible than the conventional hard quantization. In order to facilitate end-to-end training with differentiability, we propose a novel soft hyperbolic codebook quantization mechanism. Specifically, we represent the codebook $C$ as the Cartesian product of $M$ smaller codebooks, namely $C = C^1 \times C^2 \times \cdots \times C^M$, where the $m$-th codebook $C^m=\{\mathbf{c}^m_k \in \mathbb{H}^d_{\theta_m}\}_{k=1}^K$ consists of $K$ codewords distributed on the $m$-th Lorentzian manifold. Then, for the representation $\mathbf{h}_x^m $, we aim to obtain the soft quantized one $\tilde{\mathbf{h}}^m_x \in \mathbb{H}^d_{\theta_m}$ as its approximation by calculating the weighted aggregation inside the codebook $C^m$. One reasonably desirable property of $\tilde{\mathbf{h}}^m_x$ is that the expected squared distance from $\tilde{\mathbf{h}}^m_x$ to the codeword set $C^m$ should be minimum, \textit{i.e.}, 
\begin{equation}
\tilde{\mathbf{h}}^m_x \triangleq \arg \min_{\mathbf{\mu} \in \mathbb{H}^d_{\theta_m}} w^m_k \mathrm{d}^2_{\mathcal{L}_m}(\mathbf{c}^m_k, \mathbf{\mu}),
\label{fml:exp_mim}
\end{equation}
where $w^m_k$ is the computed weight of the $k$-th hyperbolic codeword $\mathbf{c}^m_k$ in the $m$-th Lorentzian manifold. Intuitively, it ensures that the quantized embedding is close to those codewords with larger weights.
It was proved in~\cite{law2019lorentzian} that with the squared Lorentzian distance  defined as:
\begin{equation}
\mathrm{d}_{\mathcal{L}_m}^2(\mathbf{a}, \mathbf{b})=\|\mathbf{a}-\mathbf{b}\|_{\mathcal{L}}^2=-2\theta_m-2\langle\mathbf{a}, \mathbf{b}\rangle_{\mathcal{L}},
\end{equation}
then Eq.~\eqref{fml:exp_mim} has the closed-form solution:
\begin{equation}
    \tilde{\mathbf{h}}^m_x = \frac{\sum^{K}_{k=1} w_k^m \mathbf{c}^m_k}{\sqrt{-1/\theta_m}   \big|  \Vert \sum^{K}_{k=1} w_k^m \mathbf{c}^m_k \Vert_{\mathcal{L}}   \big| },
\label{fml:hyper_attn}
\end{equation}
where $\left|\|\mathbf{a}\|_{\mathcal{L}}\right|=\sqrt{\left|\|\mathbf{a}\|_{\mathcal{L}}^2\right|}$ is the modulus~\cite{ratcliffe2006hyperbolic} of the imaginary Lorentzian norm of the vector $\mathbf{a}$.
Here, the attention weight $w^m_k$ is computed using softmax relaxation:
\begin{equation}
    w^m_k = \frac{\exp (- \mathrm{d}^2_{\mathcal{L}_m}(\mathbf{\mathbf{c}}^m_k, \mathbf{\mathbf{h}}^m_x) /\tau)}{\sum^{K}_j \exp (- \mathrm{d}^2_{\mathcal{L}_m}(\mathbf{c}^m_j, \mathbf{h}^m_x)/\tau)},
\end{equation}
where $\tau$ is a hyper-parameter.  After quantizing each subspace separately, we can finally get the quantized hyperbolic embedding $\tilde{\mathbf{h}}_x = (\tilde{\mathbf{h}}^1_x, \cdots, \tilde{\mathbf{h}}^M_x)$ as the approximation of original continuous embedding $\mathbf{h}_x$ on the product manifold $\mathbb{S}$. In inference, we simply use the hard quantization operation to obtain the quantized representation. For more inference details, please refer to the supplementary material.

\paragraph{Hyperbolic Quantized Contrastive Learning}
Inspired by the recent success~\cite{yan2021unsupervised,ge2023hyperbolic} of distance learning on hyperbolic space, we introduce a hyperbolic quantized contrastive learning scheme aimed at jointly both the encoder network and the quantization module. Specifically, given a mini-batch of training images $\mathcal{B}$ with the batch size $N_B$, we apply random augmentations to each image twice, resulting in total $2N_B$ views  $\{(x^{(1)}_i, x^{(2)}_i)\}_{i=1}^{N_B}$. After passing them through our encoder and the quantization module, we can obtain the quantized hyperbolic representations $\{(\tilde{\mathbf{h}}^{(1)}_{x_i}, \tilde{\mathbf{h}}^{(2)}_{x_i})\}_{i=1}^{N_B}$. Then, the hyperbolic quantized contrastive loss is defined as:
\begin{equation}
\mathcal{L}_{aug} = \frac{1}{N_B} \sum_{x \in \mathcal{B}} \left(\ell^{1}(x) + \ell^{2}(x) \right),    
\end{equation}
where $aug$ denotes ``augmented" for short, and $\ell^{j}(x)$ for $j = \{1,2\}$ is defined as:
\begin{equation}
    \ell^{(j)}(x) \!= \! - \log \frac{\mathcal{S} (\tilde{\mathbf{h}}^{(1)}_x, \tilde{\mathbf{h}}^{(2)}_x))}{\mathcal{S} (\tilde{\mathbf{h}}^{(1)}_x, \tilde{\mathbf{h}}^{(2)}_x)) + \!\!\!\! \sum\limits_{\substack{t\in {\mathcal{B}} \backslash x \\ n = 1,2}} \mathcal{S} (\tilde{\mathbf{h}}^{(1)}_x, \!\tilde{\mathbf{h}}^{(n)}_t))}.
    \label{fml:cl}
\end{equation}
Intuitively, for the hyperbolic quantized embedding $\tilde{\mathbf{h}}^{(1)}_{x}$ as the query, $\tilde{\mathbf{h}}^{(2)}_{x}$ is defined as the positive key, while other $2N_B - 2$ quantized embeddings from other images within mini-batch are considered negative keys. Different from previous works, our distance metric works on the hyperbolic product manifold $\mathbb{S}$, thus $\mathcal{S}(\tilde{\mathbf{h}}_1, \tilde{\mathbf{h}}_2)$ is defined as:
\begin{equation}
    \mathcal{S}(\tilde{\mathbf{h}}_1, \tilde{\mathbf{h}}_2) \triangleq \exp (-\mathrm{d}_{\mathbb{S}}(\tilde{\mathbf{h}}_1, \tilde{\mathbf{h}}_2)/\tau_{qc}).
    \label{fml:sim}
\end{equation}
In Eq.~\eqref{fml:sim}, the negative of the distance 
 defined on the product manifold $\mathbb{S}$ is taken as the similarity score on $S$, which implies that the similarity between two images is determined by  hyperbolic distances on $M$ decomposed Lorentzian manifold jointly; and $\tau_{qc}$ is a non-negative temperature.

\subsection{Hierarchical Semantics Learning Module}

\paragraph{Hierarchical Semantics Extraction}
Inspired by~\cite{yan2021unsupervised,guo2022hcsc}, we propose a hierarchical semantics learning module to enhance the semantics of hyperbolic quantized representation with explicit hierarchical semantic similarity as the extra training signal. To this end, we resort to  bottom-up hierarchical clustering to extract pseudo hierarchical semantic similarity. Notably, since direct clustering in the hyperbolic space is challenging, our pseudo hierarchical semantics between images is extracted based on the Euclidean tangent space $\mathcal{T}_{\mathbf{p}}\mathbb{S}$.  Our subsequent experiments showed that the semantic hierarchy refined with the Euclidean tangent space has sufficiently benefited the quality of our quantized hyperbolic representations.

Specifically, in the bottom-up merging step, the two closest sub-clusters will be merged to form a new cluster. The distance between two sub-clusters is approximated by the distance between their corresponding prototypes as:
\begin{equation}
    \mathrm{d}_{ab} = \left\| \mathbf{e}_a - \mathbf{e}_b \right\|, \ \ \mathbf{e}_a = \frac{1}{n_a}\sum_{\mathbf{v}_j \in G_a} \mathbf{v}_j,
\end{equation}
where $\left\|\cdot \right\|$ denotes the Euclidean distance, $n_a $ represents the number of samples belonging to the sub-cluster $G_a$, and the prototype $\mathbf{e}_a$ is the average of all tangent vectors in $G_a$.

In practice, the distance threshold for merging can be set as sufficiently high, allowing the merging process to continue regardless of the distance of the closest sub-cluster pair. Then, we can pre-define $L$ levels, with each level containing $N_l$ clusters. When the bottom-up merging reaches a cluster count of $N_l$, we save the corresponding clustering result. Ultimately, a hierarchical similarity structure with $L$ levels is obtained. After completing the hierarchical clustering, as the obtained prototypes are in the tangent space, we need to map them to the product manifold $\mathbb{S}$ so that they can interact with our quantized hyperbolic representations. For this purpose, the prototype $\mathbf{e}_j = (\mathbf{e}_j^1, \cdots, \mathbf{e}_j^M)$ is transformed to the product manifold via the exponential map:
\begin{equation}
    \tilde{\mathbf{e}}_j = \exp_{\mathbf{p}}(\mathbf{e}_j) = \left(\exp_{\mathbf{p}^1}^{\theta_1}(\mathbf{e}^1_j), \cdots, \exp_{\mathbf{p}^M}^{\theta_M}(\mathbf{e}^M_j) \right).
\end{equation} 
Practically, directly performing hierarchical clustering on all training images is time-consuming. Therefore, during the hierarchical clustering phase, we first apply K-means clustering to the training data to obtain a sufficient number of  sub-clusters, and then perform iterative merging to get the desired $L$ hierarchies. Moreover, we perform hierarchical clustering at the beginning of every epoch for training efficiency.

\begin{figure}
    \centering
    \includegraphics[width=0.78\linewidth]{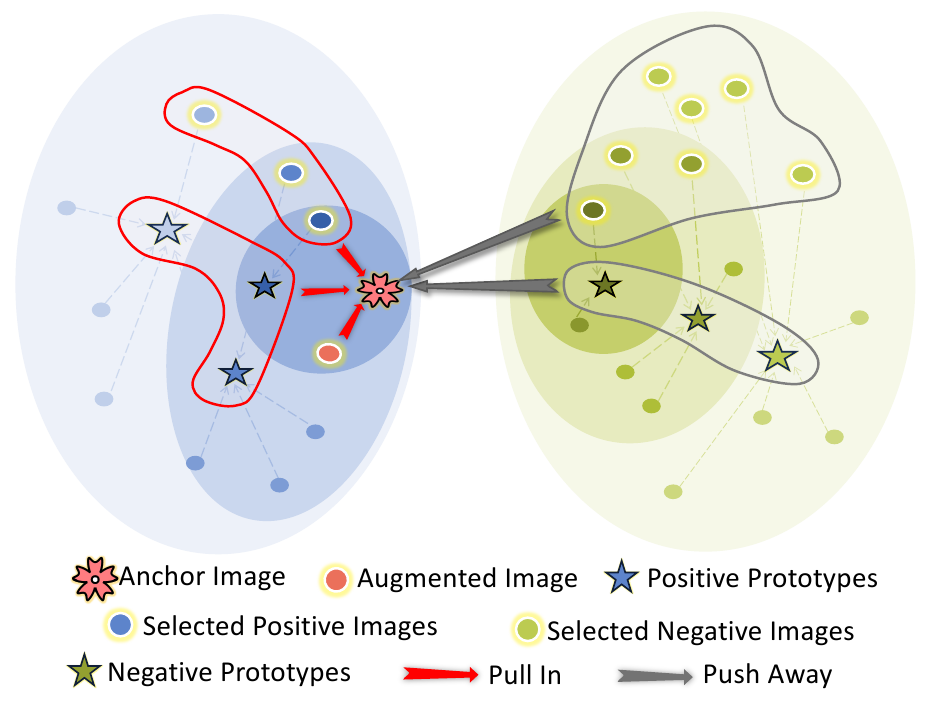}
    \vspace{-2mm}
    \caption{Illustration of both instance-wise and prototype-wise contrastive learning based on our extracted hierarchy.}
    \vspace{-6mm}
    \label{fig:hcl}
\end{figure}

\paragraph{Prototype-Wise \& Instance-Wise Semantics Learning}
The multi-level hierarchical semantics relationship we uncover consists of two components: the affiliation relationship between images and their corresponding prototypes, and the assignment relationship between images within the semantics structure. Therefore, we resort to both prototype-wise~\cite{li2020prototypical,guo2022hcsc} and instance-wise contrastive loss to jointly learn these two aspects. Specifically, given a query image $x_i$, we denote the hyperbolic prototype $\tilde{\mathbf{e}}^l_j(i)$ that $x_i$ is assigned to in $l$-th hierarchy and regard $(\tilde{\mathbf{h}}_{x_i}, \tilde{\mathbf{e}}^l_j(i))$ as a positive pair, while other prototypes at the same level in the hierarchy play the role of negative prototypes. Then the prototype-wise contrastive loss is defined as:
\begin{equation}
    \mathcal{L}_{prot} = - \sum^{N_B}_{i=1} \frac{1}{L} \sum^{L}_{l=1} \log
    \frac{\mathcal{S}(\tilde{\mathbf{h}}_{x_i}, \tilde{\mathbf{e}}^l_j(i))}{\sum_{n=1}^{N_l} \mathcal{S}(\tilde{\mathbf{h}}_{x_i}, \tilde{\mathbf{e}}^l_n) },
    \label{fml:hpl}
\end{equation}
where the similarity metric $\mathcal{S(\cdot, \cdot)}$ is the negative hyperbolic distance defined in Eq.~\eqref{fml:sim}. 

While the prototypical contrastive loss in Eq.~\eqref{fml:hpl} can preserve similarity between the image and their closest prototype in a multi-granularity hierarchy, it neglects the proximity between images belonging to the same prototype in each hierarchy. To alleviate this issue, we fall back on instance-wise contrastive learning. Specifically, given an image $x_i$, we randomly select an image $x_{il}$ from the pool of images assigned to the prototypes that $x_i$ belongs to at each $l$-th hierarchy. Then the hierarchical instance-wise contrastive loss is defined as:
\begin{equation}
    \mathcal{L}_{ins} = - \sum^{N_B}_{i=1} \frac{1}{L} \sum^{L}_{l=1} \log \frac{\mathcal{S} (\tilde{\mathbf{h}}_{x_i}, \tilde{\mathbf{h}}_{x_{il}})) }{\sum\limits_{\substack{t\in {\mathcal{B}} \backslash x_i}} \mathcal{S} (\tilde{\mathbf{h}}_{x_i}, \tilde{\mathbf{h}}_t))}.
    \label{fml:redefine_cl}
\end{equation}

Then, our training objective for the pseudo hierarchical similarity is defined as:
\begin{equation}
    \mathcal{L}_{hs} = \lambda_1 \mathcal{L}_{prot} +  \lambda_2 \mathcal{L}_{ins},
    \label{fml:hs}
\end{equation}
where $\lambda_1$ and $\lambda_2$ are the hyper-parameters used to control the relative importance of the two losses. Figure~\ref{fig:hcl} visualizes our hierarchical semantics learning framework. Overall, the ultimate training objective for HiHPQ is:
\begin{equation}
    \mathcal{L} = \mathcal{L}_{aug} + \mathcal{L}_{hs}.
\end{equation}

\begin{table*}[!t]
\centering
\begin{tabularx}{\textwidth}{l c XXX XXX XXX XXX}
\toprule
\multirow{2}{*}{Method}& \multirow{2}{*}{Type} & \multicolumn{3}{c}{Flickr25K} & \multicolumn{3}{c}{CIFAR-10 (I)} & \multicolumn{3}{c}{CIFAR-10 (II)} & \multicolumn{3}{c}{NUS-WIDE} \\

 &  & 16bits & 32bits & 64bits & 16bits & 32bits & 64bits & 16bits & 32bits & 64bits & 16bits & 32bits & 64bits \\
\midrule

LSH+VGG &BH     &56.11  &57.08  &59.26  &14.38  &15.86  &18.09  &12.55  &13.76  &15.07  &38.52  &41.43  &43.89 \\ 
SH+VGG  &BH     &59.77  &61.36  &64.08  &27.09  &29.44  &32.65  &27.20  &28.50  &30.00  &51.70  &51.10  &51.00 \\ 
SpH+VGG &BH     &61.32  &62.47  &64.49  &26.90  &31.75  &35.25  &25.40  &29.10  &33.30  &49.50  &55.80  &58.20 \\ 
ITQ+VGG &BH     &63.30  &65.92  &68.86  &34.41  &35.41  &38.82  &30.50  &32.50  &34.90  &62.70  &64.50  &66.40 \\ 
DeepBit &BH     &62.04  &66.54  &68.34  &19.43  &24.86  &27.73  &20.60  &28.23  &31.30  &39.20  &40.30  &42.90 \\ 
SGH     &BH     &72.10  &72.84  &72.83  &34.51  &37.04  &38.93  &43.50  &43.70  &43.30  &59.30  &59.00  &60.70 \\ 
HashGAN &BH     &72.11  &73.25  &75.46  &44.70  &46.30  &48.10  &42.81  &47.54  &47.29  &68.44  &70.56  &71.71 \\ 
GreedyHash  &BH &69.97  &70.85  &73.03  &44.80  &47.20  &50.10  &45.76  &48.26  &53.34  &63.30  &69.10  &73.10 \\ 
BinGAN  &BH     &-      &-      &-      &-      &-      &-      &47.60  &51.20  &52.00  &65.40  &70.90  &71.30 \\ 
BGAN    &BH     &-      &-      &-      &-      &-      &-      &52.50  &53.10  &56.20  &68.40  &71.40  &73.00 \\ 
SSDH    &BH     &75.65  &77.10  &76.68  &36.16  &40.17  &44.00  &33.30  &38.29  &40.81  &58.00  &59.30  &61.00 \\ 
DVB     &BH     &-      &-      &-      &-      &-      &-      &40.30  &42.20  &44.60  &60.40  &63.20  &66.50 \\ 
TBH     &BH     &74.38  &76.14  &77.87  &54.68  &58.63  &62.47  &53.20  &57.30  &57.80  &71.70  &72.50  &73.50 \\ 
Bi-half Net &BH    &76.07  &77.93  &78.62  &56.10  &57.60  &59.50  &49.97  &52.04  &55.35  &76.86  &78.31  &79.94 \\ 
CIBHash &BH     &77.21  &78.43  &79.59  &59.39  &63.67  &65.16  &59.00  &62.20  &64.10  &79.00  &80.70  &81.50 \\ 
\midrule
PQ+VGG  &PQ     &62.75  &66.63  &69.40  &27.14  &33.30  &37.67  &28.16  &30.24  &30.61  &65.39  &67.41  &68.56 \\ 
OPQ+VGG &PQ     &63.27  &68.01  &69.86  &27.29  &35.17  &38.48  &32.17  &33.50  &34.46  &65.74  &68.38  &69.12 \\ 
DeepQuan &PQ    &-      &-      &-      &39.95  &41.25  &43.26  &-      &-      &-      &-      &-      &-      \\ 
SPQ  &PQ     &77.35  &78.74  &79.98  &63.17  &66.88  &68.02  &56.56  &61.45  &63.30  &78.51  &80.41  &81.70      \\ 
MeCoQ   &PQ     &\textbf{81.31}  &81.71  &82.68  &68.20  &69.74  &71.06  &62.88  &64.09  &65.07  &78.31  &81.59  &81.80 \\ 
HiHPQ    &PQ     &80.74  &\textbf{82.55}  &\textbf{82.95}      &\textbf{70.56}  &\textbf{73.22}  &\textbf{73.71}  &\textbf{63.31}  &\textbf{65.81}  &\textbf{67.11}  &\textbf{79.86}      &\textbf{82.05}      &\textbf{82.63} \\
\bottomrule
\end{tabularx}
\vspace{-1.5mm}
\caption{MAP ($\%$) comparision between different unsupervised state-of-the-art efficient retrieval methods on benchmark datasets.  The types of methods used are denoted as ``PQ" for ``Product Quantization" and ``BH" for ``Binary Hashing".}
\vspace{-3mm}
\label{tbl: main}
\end{table*}

\vspace{-1.8mm}
\section{Experiments}
\subsection{Experimental Setup}

\paragraph{Datasets} 
The proposed method is evaluated using \textbf{Flickr25K}~\cite{huiskes2008mir}, \textbf{NUS-WIDE}~\cite{chua2009nus}, as well as two experimental protocols CIFAR-10 (I) and CIFAR-10 (II) both of which are based on \textbf{CIFAR-10}~\cite{krizhevsky2009learning}. Further details can be found in the supplementary material.

\paragraph{Evaluation Metrics} We utilize the Mean Average Precision (MAP) at top N as the evaluation metric to assess the proposed method. We adopt the MAP@1000 for CIFAR-10(I) and CIFAR-10 (I) and (II), and MAP@5000 for both NUS-WIDE and Flickr25K.

\paragraph{Baselines} We consider following unsupervised baselines for comparison: (i) binary hashing methods: LSH~\cite{charikar2002similarity}, SH~\cite{weiss2008spectral}, SpH~\cite{heo2012spherical}, ITQ~\cite{gong2012iterative};  DeepBit~\cite{lin2016learning}, SGH~\cite{dai2017stochastic}, HashGAN~\cite{dizaji2018unsupervised}, GreedyHash~\cite{su2018greedy}, BinGAN~\cite{zieba2018bingan}, BGAN~\cite{song2018binary}, SSDH~\cite{yang2018semantic}, DVB~\cite{shen2019unsupervised}, TBH~\cite{shen2020auto}, Bi-half Net~\cite{li2021deep}, CIBHash~\cite{QiuSOYC21}; and 
(ii) product quantization methods: PQ~\cite{jegou2010product}, OPQ~\cite{ge2013optimized}; DeepQuan~\cite{chen2018learning}, SPQ\footnote{As SPQ uses a totally different backbone and training strategy, we reproduce SPQ in our setting using their codebase.}~\cite{jang2021self}, MeCoQ~\cite{wang2022contrastive}.

\subsection{Implementation Details} 
Following~\cite{wang2022contrastive}, the encoder is constituted with the pre-trained VGG-16 network followed by a linear projector. Our model is implemented using Pytorch~\cite{paszke2019pytorch}. Riemannian SGD~\cite{bonnabel2013stochastic} is utilized for optimization. Regarding hyper-parameters for the hyperbolic product quantizer, the number of codewords $K$ in each codebook $C^m$ is fixed  to 256, and the dimension of each codebook is fixed to 16. By setting the number of small codebooks $M$ as $\{2,4,8\}$, the final codeword in the codebook $C$ is represented by $\{16,32,64\}$ bits according to $B= M \log_2K$.  The pre-defined hierarchies for different datasets are: $[50,20]$ on Flickr25K; $[200,100,50]$ on CIFAR-10 (I); $[100,50,25 ]$ on CIFAR-10 (II); $[200, 100, 75]$ on NUS-WIDE. Please refer to the supplementary material for more implementation details.

\begin{figure*}[!t]
\centering
\subfigure[]{
\includegraphics[width=0.23\linewidth]{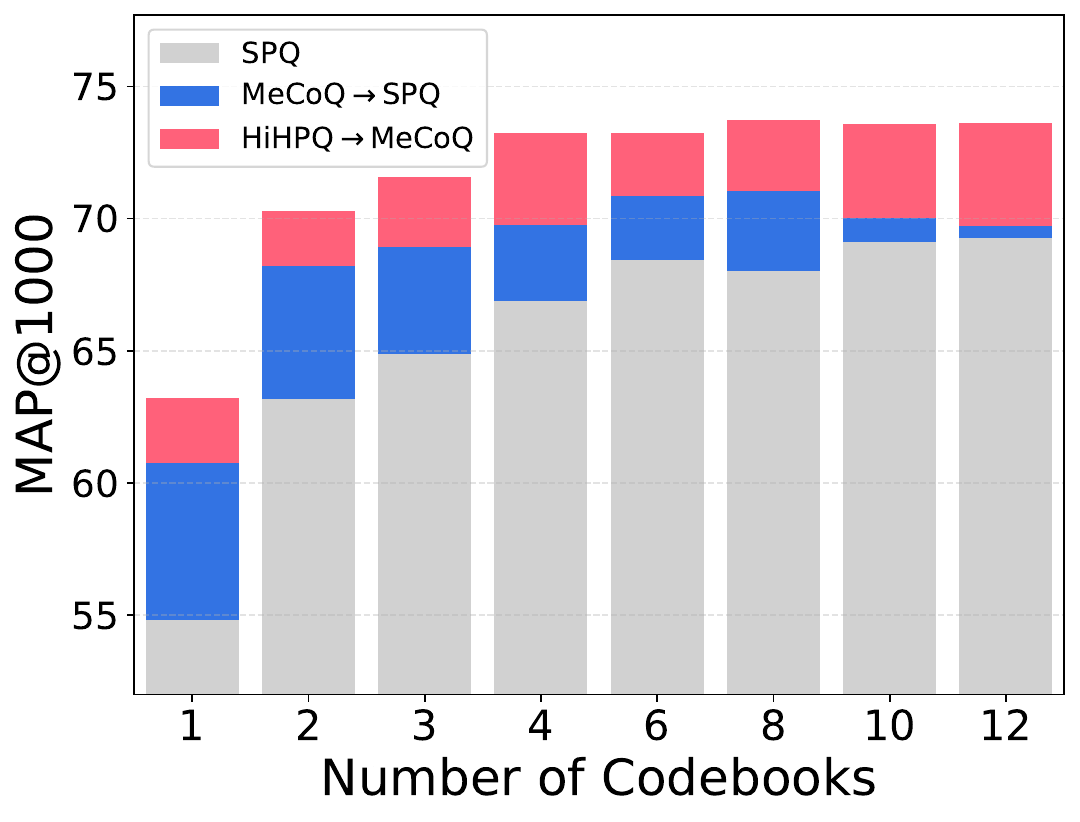}
\label{fig:codebook}
}
\subfigure[]{
\includegraphics[width=0.23\linewidth]{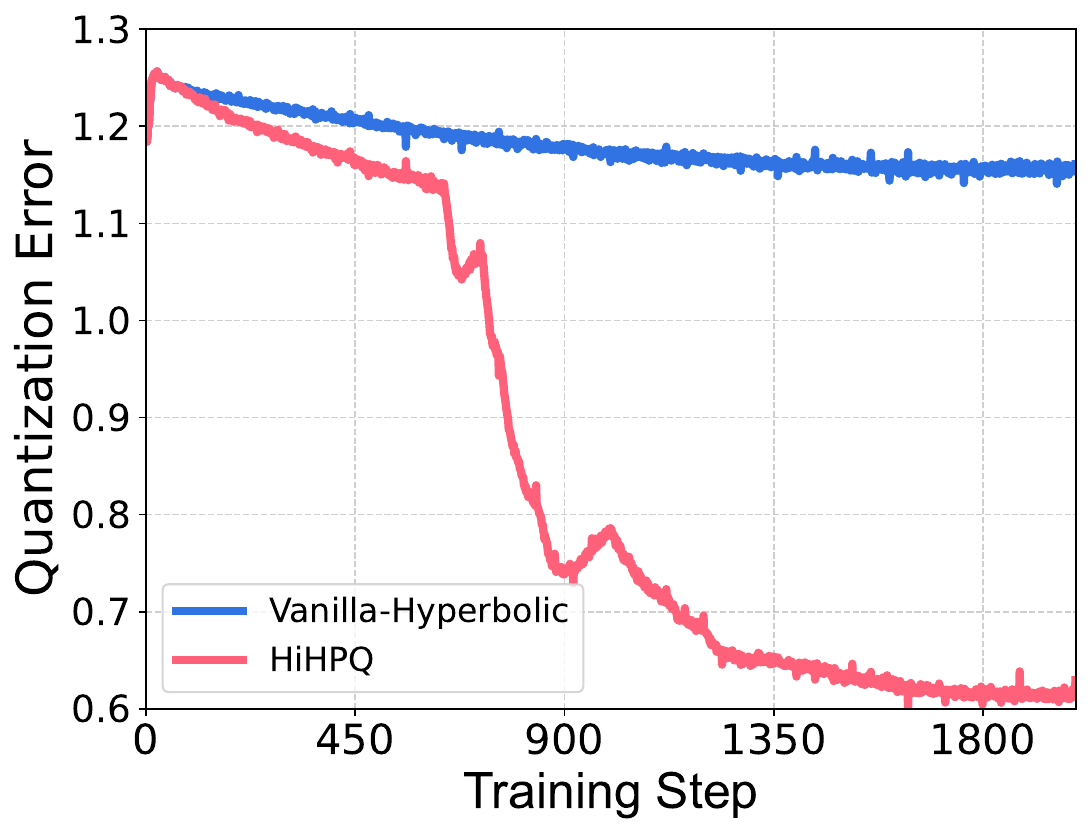}
\label{fig:quant_error}
} 
\vspace{-0.2cm}
\subfigure[]{
\includegraphics[width=0.23\linewidth]{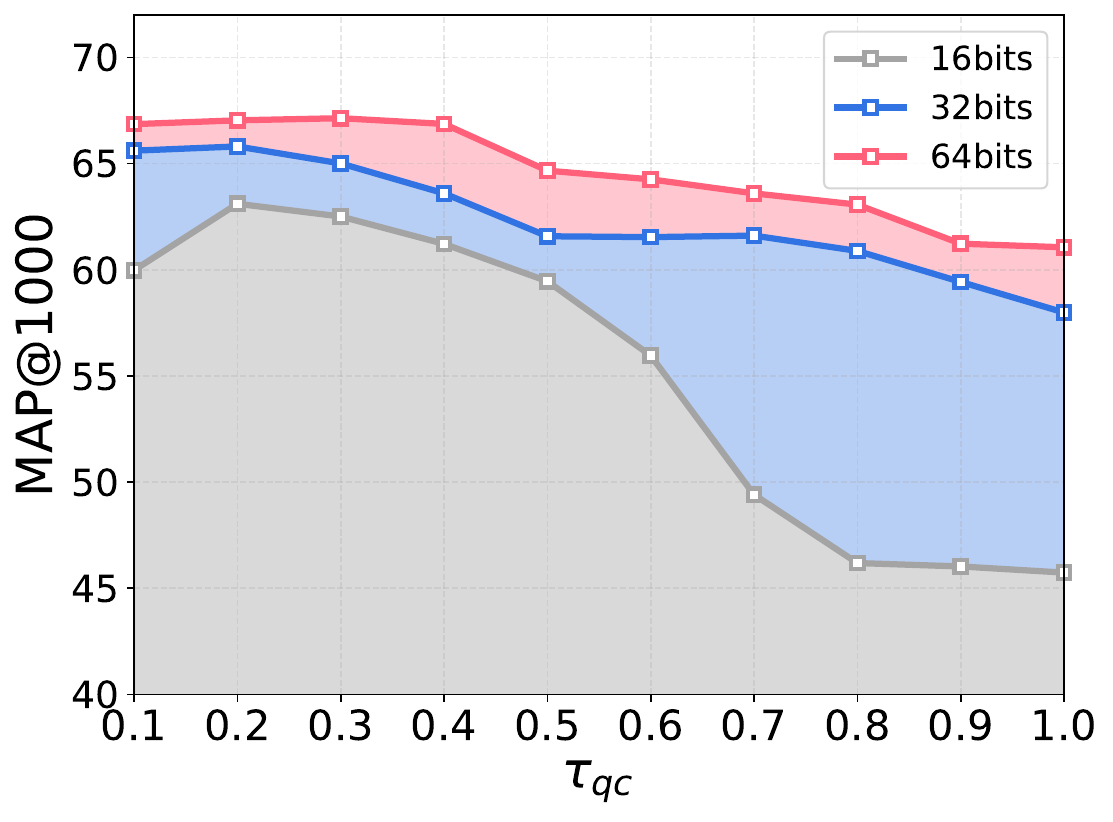}
\label{fig:temp}
}
\subfigure[]{
\includegraphics[width=0.235\linewidth]{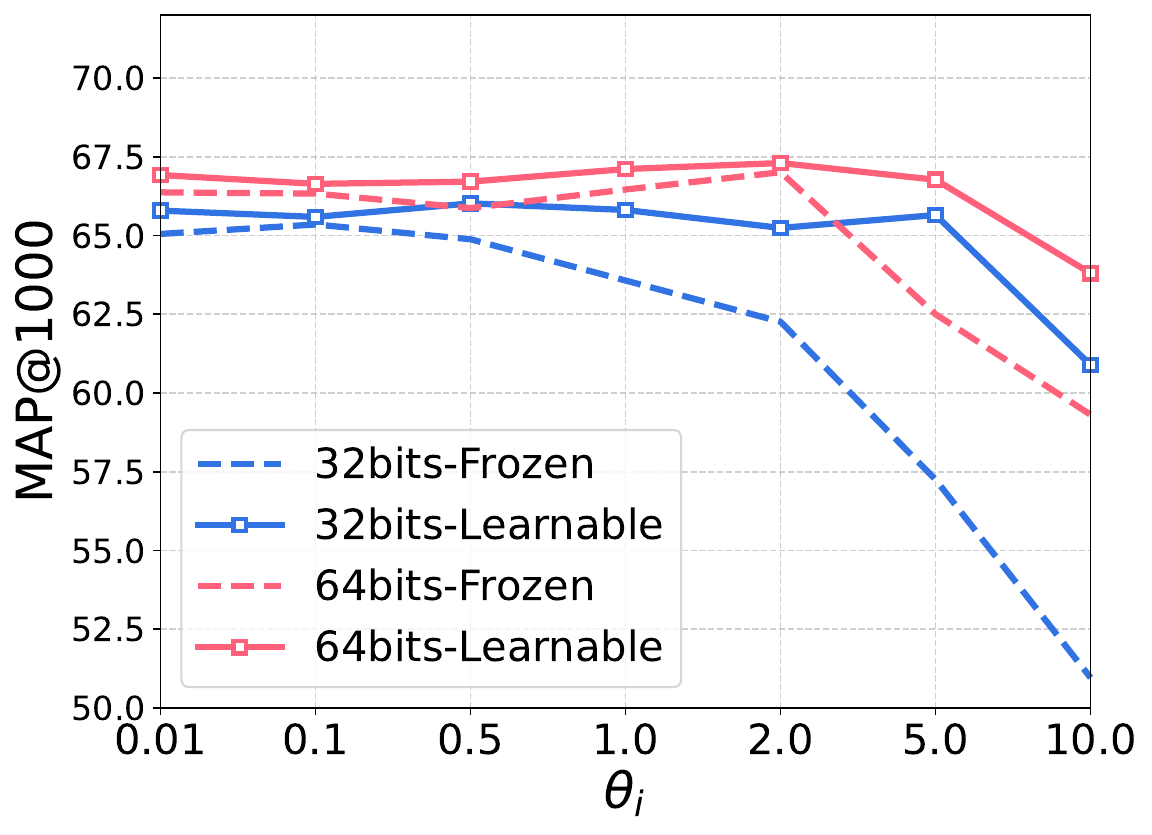}
\label{fig:curv}
}
\vspace{-1.5mm}
\caption{(\textbf{a}) MAP@1000 with varying numbers of codebooks on CIFAR10 (I); ``A $\rightarrow$ B" denotes the performance gain achieved by Model A compared to Model B. (\textbf{b}) 32-bit quantization errors of HiHPQ and its variant on CIFAR10 (II).
(\textbf{c}) Effects of the temperature $\tau_{qc}$ on CIFAR-10 (II). (\textbf{d}) Effect of the initial curvature parameters on CIFAR10 (II). }
\vspace{-2.5mm}
\end{figure*}

\subsection{Results and Analyses}
\paragraph{Overall Performance} Table \ref{tbl: main} presents the performances of our proposed model and existing baselines on public benchmark image datasets. It can be seen that our proposed HiHPQ outperforms previous shallow and deep quantization methods by a significant margin across almost all experimental settings. Among all the baselines compared, our method closely resembles SPQ and MeCoQ, which propose cross-quantized contrastive learning and debiased contrastive learning with code memory, respectively. However, these two methods significantly underperform HiHPQ, which can be attributed to HiHPQ's hierarchical semantics learning in the hyperbolic product manifold. In Figure~\ref{fig:codebook}, we further demonstrate the retrieval performance of HiHPQ with varying numbers of codebooks. It can be observed that regardless of the number of codebooks, our model outperforms the state-of-the-art SPQ and MeCoQ by a significant margin, showing the superiority of the HiHPQ.

\paragraph{Ablation Study}

We configure six variants of HiHPQ for investigation. (i) \textit{Vanilla-Hyperbolic}: by setting both $\lambda_1=0$ and $\lambda_2=0$, it discards the extracted hierarchy as the extra training signal. (ii)  \textit{Instance-Hyperbolic}: by setting $\lambda_1=0$ and $\lambda_2=1$, it learns the hierarchical similarity by only using the instance-wise contrastive loss in Eq.~\eqref{fml:redefine_cl}. (iii) \textit{Prototype-Hyperbolic}: by setting $\lambda_1=1$ and $\lambda_2=0$, it learns the hierarchical similarity by only using the prototype-wise contrastive loss in Eq.~\eqref{fml:hpl}.  To compare the representative power of hyperbolic space with Euclidean space, we implement the Euclidean counterparts of above three mentioned variants: (iv) \textit{vanilla-Euclidean}, (v) \textit{Instance-Euclidean}, and (vi) \textit{Prototype-Euclidean}. For the Euclidean scenario, the quantization process and the distance metric refer to MeCoQ~\cite{wang2022contrastive}, and the continuous features ahead of the product quantization module are utilized to construct the hierarchical similarity.

\begin{table}[!t]
    \centering
    \begin{tabularx}{0.9\linewidth}{lccc}
    \toprule
    \textbf{Variants}            &16bits       &32bits        &64bits    \\
    \midrule
    Vanilla-Euclidean            &59.26        &62.52         &62.72      \\
    Vanilla-Hyperbolic           &58.40        &61.86         &62.17     \\
    Instance-Euclidean           &61.56        &64.70         &65.40      \\
    Instance-Hyperbolic          &61.80        &63.84         &63.98     \\
    Prototype-Euclidean          &60.84        &64.36         &65.10     \\
    Prototype-Hyperbolic         &62.65        &65.54         &66.85     \\ 
    \midrule
    HiHPQ (Full Model)            &\textbf{63.31}        &\textbf{65.81}         &\textbf{67.11}      \\

    \bottomrule
    \end{tabularx}
    \caption{MAP ($\%$) comparision between HiHPQ and its variants on CIFAR10 (II).}
    \label{tab: ablation}
    \vspace{-4mm}
\end{table}
As seen from Table~\ref{tab: ablation}, when employing the vanilla view-augmented contrastive objective, quantized representations in  hyperbolic space may not have necessarily better quality than the counterpart in Euclidean space. Moreover, by introducing the extracted hierarchical similarity as the extra training signal, the retrieval performance of both embedding spaces can be significantly improved. Nevertheless, if only using the instance-wise contrastive loss to learn the hierarchical similarity between images, quantized representations in Euclidean space achieve greater performance gain than the hyperbolic scenario. It is worth noting that the enhancement in performance is particularly more significant when employing the prototype-wise contrastive loss on hyperbolic space to acquire hierarchical similarity. We conjecture this is because the prototype-wise objective directly incorporates prototypes as internodes in a tree-like hierarchy, thereby leveraging the advantages of hyperbolic geometry.

In Figure~\ref{fig:quant_error}, we illustrate the effect of hierarchical semantics on the quantization error. Here, the quantization error refers to the hyperbolic distance between the continuous representation of an image and its quantized representation. It is shown that including hierarchical information as extra training supervision leads to a significant reduction in quantization error, while its absence results in a slower rate of decrease in error.

\paragraph{Impacts of Pre-defined Clustering Sizes}
We investigate the impact of different pre-defined hierarchies on retrieval performance. As seen from Table \ref{tab:hierarchy_size}, setting a cluster number for each hierarchy level that is significantly larger than the number of ground truth labels in the dataset (\textit{e.g.}, 200 for CIFAR10 (I)) often has a more positive impact on performance. Moreover, setting the hierarchy level to 2 or 3 generally yields great performance improvements. In particular, we observe that prototype-wise contrastive loss in Eq.~\eqref{fml:hpl} always outperforms the instance-wise one in Eq.~\eqref{fml:redefine_cl} in terms of performance gain, which is consistent with the previous conclusion of the ablation study.

\paragraph{Hyper-parameter Sensitivity Analysis}
\begin{table}[!t]
    \centering
    \begin{tabular}{c|ccc}
        \toprule
         Dataset                        &Hierarchy Size &Prototype  &Instance \\
         \midrule
         \multirow{6}{*}{CIFAR10 (I)}           &\multicolumn{3}{c}{Vanilla-Hyperbolic $\rightarrow$ 66.87}      \\
         \cline{2-4}
         \noalign{\smallskip}
         ~                              &200            &72.12      &71.54      \\
         ~                              &100            &71.85      &72.01      \\
         ~                              &200,100        &72.97      &72.49      \\
         ~                              &200,100,50     &73.58      &72.06      \\
         ~                              &100,50,25      &71.72      &70.39      \\
         \midrule
         \multirow{6}{*}{Flickr25K}     &\multicolumn{3}{c}{Vanilla-Hyperbolic $\rightarrow$ 80.63}      \\
         \cline{2-4}
         \noalign{\smallskip}
         ~                              &50          &81.35      &81.21      \\
         ~                              &100,50      &81.79      &81.55      \\ 
         ~                              &100,30      &81.99      &82.33      \\
         ~                              &50,20       &82.63      &81.93      \\
         ~                              &100,50,20   &81.91      &81.71      \\
         \bottomrule
    \end{tabular}
    \vspace{-1mm}
    \caption{MAP (\%) performance on CIFAR10 (I) at 32bits using different preset hierarchical structures. The ``Prototype" and ``Instance" columns show the performances of previously mentioned variants ``Prototype-Hyperbolic" and ``Instance-Hyperbolic", respectively.}
    \label{tab:hierarchy_size}
    \vspace{-3.5mm}
\end{table}

We analyze the effect of two key hyper-parameters: the temperature $\tau_{qc}$ in contrastive learning, and the curvature parameters $\theta_i$. As shown in Figure~\ref{fig:temp}, a typically small temperature value (\textit{e.g.}, 0.2) has a positive impact on retrieval performance. Figure~\ref{fig:curv} illustrates the influence of the initial value of $\theta_i$ on model performance, indicating that our method is robust within the range $(0.01, 2.0)$ of initial values, whereas larger values lead to degradation. Notably, setting $\theta_i$ as a learnable parameter can alleviate this decline in model performance.

\section{Conclusion}
In this paper, we have proposed a novel unsupervised deep product quantization method, namely HiHPQ. In HiHPQ, we managed to build a hyperbolic product quantizer, in which both the soft hyperbolic codebook quantization mechanism and the quantized contrastive learning based on the hyperbolic product manifold were introduced to facilitate quantization. We further enhanced the semantics of quantized representations by utilizing the extracted hierarchical semantics supervision in our semantics learning module, which helped better distinguish between the query's similar images and non-matching ones. Extensive experiments have shown the superiority of our model over existing baselines.

\section{Acknowledgements}
The work described in this paper was substantially supported by InnoHK initiative, The Government of the HKSAR, and Laboratory for AI-Powered Financial Technologies. In addition, the work was partially supported by a grant from the Research Grants Council of the Hong Kong Special Administrative Region, China (CUHK 14222922, RGC GRF No. 2151185).

\bibliography{references}

\appendix
\section{Appendix}
\paragraph{Encoding and Retrieval}
For testing, we encode all images in the database using the hard quantization operation to obtain the quantized hyperbolic embedding $\tilde{\mathbf{h}}_{x_d} = (\tilde{\mathbf{h}}_{x_d}^1, \cdots, \tilde{\mathbf{h}}_{x_d}^M)$ as:
\begin{equation}
    \tilde{b}^m_{x_d} = \underset{k}{\arg \min \ }\mathrm{d}_{\mathcal{L}_m}(\mathbf{c}^m_k, \mathbf{h}_{x_d}^m), \ \ \ \tilde{\mathbf{h}}_{x_d}^m = C^m[\tilde{b}^m_{x_d}].
\end{equation}
To recover $\tilde{\mathbf{h}}_{x_d}$ for the image $x_d$, we only need to store $M$ codeword indices $\{\tilde{b}^m_{x_d}\}_{m=1}^M$ it is assigned to, which only needs storage consumption of $M \log_2 K$ bits. During the retrieval phase, given a query image embedding $\mathbf{h}_{x_q} = (\mathbf{h}_{x_q}^1, \cdots, \mathbf{h}_{x_q}^M)$,  we first pre-compute a query-specific look-up table $ \mathbf{T_q}\in \mathbb{R}^{M \times k}$ storing the Lorentzian distance between the query and all codewords, of which each element $\mathbf{T_q}[m,k] = \mathrm{d}_{\mathcal{L}_m}(\mathbf{h}_{x_q}^m, \mathbf{c}^m_k)$. Finally, the distance between the query $x_q$ and the database point $x_d$ can be efficiently computed by summing up the chosen values from the look-up table as:
\begin{equation}
    \mathrm{d}_{\mathbb{S}}(\mathbf{h}_{x_q}, \tilde{\mathbf{\mathbf{h}}}_{x_d}) = \sum^M_{m=1}\mathbf{T_q}[m, \tilde{b}^m_{x_d}].
\end{equation}

\paragraph{Datasets}
Three datasets are adopted to evaluate the performance of the proposed product quantization method. \textbf{Flickr25K}~\cite{huiskes2008mir} comprises 25,000 images grouped into 24 categories. In our setting, 2,000 images are randomly selected as testing queries while 5,000 images are randomly chosen as the training set from the remaining images. \textbf{CIFAR-10}~\cite{krizhevsky2009learning} is a dataset of 60,000 images divided into 10 categories. For evaluation, two experiment protocols are employed. In CIFAR-10 (I), a test query set containing 1,000 images per category (10,000 images in total) is used, while the remaining 50,000 images are utilized for training. In CIFAR-10 (II), 1,000 images per category are randomly selected as testing queries, with 500 per category serving as the training set. In both protocols, the retrieval database included all images except those in the test query set. \textbf{NUS-WIDE}~\cite{chua2009nus} contains roughly 270k images that are divided into 81 categories. For evaluation purposes, a subset of the 21 most popular categories is selected. The testing queries consist of 100 randomly selected images per category, while the remaining images make up the database. Furthermore, 500 images per category from the database are sampled, forming a training set containing 10,500 images.
\paragraph{Implementation Details}We employ the raw image pixels as input and employ the pre-trained VGG16 model, specifically its conv1 to fc7 layers, as the backbone network. The initial learning rate was set to 1e-3 and decreased in a cosine decay manner until it reached 1e-5. We utilize the data augmentation scheme in~\cite{wang2022contrastive} comprising random cropping, horizontal flipping, image graying, random color distortions, and Gaussian blurs. During training, the batch size is set as 64, and the maximum epoch is set as 50.
the curvature parameter $-\theta_i$ in each Lorentzian manifold is initialized as $1.0$ and is designated as learnable. Following \cite{chen2021fully,guo2022clipped}, we restrict to the norm of the last $d$ dimension of embeddings in each Lorentzian manifold to be no bigger than $1.5$, so as to prevent numerical instability. The default temperature $\tau$ of the hyperbolic codebook attention mechanism is set as $0.2$, while the temperature $\tau_{qc}$ of contrastive learning is chosen from $\{0.1, 0.2, 0.3\}$. The loss weights $\lambda_1$ and $\lambda_2$  are set as $1.0$ and $0.1$ respectively. 

As for the hyperparameters regarding the quantization module, It is worth noting that we set the dimension of each codeword in the codebook as 16. In actuality, this is describing a 15-dimension hyperbolic space characterized by the Lorentz model.

\paragraph{Discussion on Experimental Setting}
SPQ~\cite{jang2021self} used an initialized ResNet50 as the backbone in their paper and trained it for thousands of epochs on each dataset. Their configuration differed from the proposed model and the majority of baseline models. Therefore, we replicated their results using their official codebase under our experimental settings (pre-trained VGG16 network as the backbone, maximum epoch as 50). Additionally, we discovered that when training on NUS-WIDE, MeCoQ~\cite{wang2022contrastive} used the entire database as its training set (193,734 images), while the majority of compared baselines used 10,500 training images. As a result, we adjusted the training set of MeCoQ on NUS-WIDE to the 10,500 sampled images and re-ran the dataset based on their official codebase, subsequently reporting the corresponding performance.

\clearpage

\end{document}